\documentclass[a4paper]{jpconf}
\usepackage{graphicx}
\usepackage{amsmath}
\usepackage{color}
\begin{document}
\title{Application of the parallel multicanonical method to lattice gas condensation}

\author{Johannes Zierenberg, Micha Wiedenmann and Wolfhard Janke}

\address{Institut f\"ur Theoretische Physik, Universit\"at Leipzig,
         Postfach 100\,920, D-04009 Leipzig, Germany}

\ead{zierenberg@itp.uni-leipzig.de}

\begin{abstract}
We present the speedup from a novel parallel implementation of the multicanonical method on the example 
of a lattice gas in two and three dimensions. In this approach, all cores perform independent equilibrium
runs with identical weights, collecting their sampled histograms after each iteration in order to estimate
consecutive weights. The weights are then redistributed to all cores. These steps are repeated until 
the weights are converged. This procedure benefits from a minimum of communication while distributing the
necessary amount of statistics efficiently. 

Using this method allows us to study a broad temperature range for a variety of large and complex systems.
Here, a gas is modeled as particles on the lattice, which interact only with their nearest neighbors.
For a fixed density this model is equivalent to the Ising model with fixed magnetization. We compare our 
results to an analytic prediction for equilibrium droplet formation, confirming that a single macroscopic
droplet forms only above a critical density.
\end{abstract}

\section{Introduction}
Despite the large amount of existing literature, condensation remains under investigation up to today
by means of analytical treatment, numerical studies, and experiments
~\cite{Binder1, Pleimling1, Biskup1, Biskup2, Neuhaus1, Binder2, Nussbaumer1, Nussbaumer2}.
In addition to the increasing interest in non-equilibrium properties and droplet size distributions, questions
regarding the equilibrium properties remain open as well.
For example, Biskup et al.~\cite{Biskup1, Biskup2} showed that in equilibrium an over-saturated gas system 
is either in the evaporated phase or in the condensed phase consisting of a single macroscopic droplet. 
The probability for droplets of intermediate size is negligibly small.
This result was proven for the two-dimensional Ising model and numerically verified~\cite{Nussbaumer1,Nussbaumer2}.
We set out to test this analytical result also for the three-dimensional case of a lattice gas.
Due to the nucleation barrier, we applied a parallel implementation~\cite{PMUCA} (for related 
methods see \cite{PMUCA2,PMUCA3,PMUCA4}) of multicanonical simulations~\cite{MUCA1,MUCA2,MUCA3} 
to circumvent a possible hysteresis, driving the system back and forth between the condensed 
and evaporated phase.

We start by describing the main method and the theoretical arguments together with the model.
This is followed by a comparison to micromagnetic simulations of the equivalent Ising model in order
to verify the correctness. We discuss the speedup with increasing parallelization and compare the 
simulation effort with the independent micromagnetic simulation.
Afterwards, the results in two and three dimensions are presented and discussed, followed by
some concluding remarks.

\section{Method}
The multicanonical method \cite{MUCA1,MUCA2,MUCA3} enables the sampling of a broad parameter 
space in a single simulation by modifying the acceptance probability according to a simple idea.
The Boltzmann factor is replaced, or sometimes extended, by a weight function with the purpose to
increase the probability to reach states that are suppressed otherwise. This can be achieved for a variety
of order parameters leading to multimagnetic, multibondic and many more realizations but can best be explained
for the case of the internal energy $E$. Formally, this can be described by writing the partition function in
terms of the density of states $\Omega(E)$,
\begin{equation}
  Z_{\rm can} = \sum_{E}\Omega(E)e^{-\beta E} \rightarrow \sum_{E}\Omega(E)W(E),
\end{equation}
where we have already transformed the Boltzmann factor into the weight function $W(E)$. Analogous to the
Metropolis algorithm, new states ($E_n$) are generated from old states ($E_o$) and accepted with 
\begin{equation}
  \min \left(1,\frac{W(E_n)}{W(E_o)}\right). \nonumber
\end{equation}
If we were to know $\Omega(E)$, we could sample the full energy range with a flat distribution choosing 
\mbox{$W(E)=1/\Omega(E)$}.
However, this is in general not the case such that $W(E)$ has to be obtained in an iterative way. 
This can be simple or complex; it can modify the weights on the fly or rely on equilibrium distributions
in each iteration. 
A simple example would be to perform equilibrium simulations with initial weights $W^{i}$ and estimate the 
consecutive weights as $W^{i+1}(E) = W^{i}(E)/H^i(E)$. Hence, if a state occurs more often the weight is 
reduced more strongly.

We applied a parallel version of this multicanonical method which has been shown to exhibit an ideal scaling
for the Ising model and a very good scaling for the $q=8$ Potts model~\cite{PMUCA}. 
This parallel implementation relies on equilibrium iterations, distributing the sampling of the
distribution $H^{i}(E)$ on $p$ independent processes. Since the weights are not modified within a single iteration,
each process returns a contribution to the same distribution and they can simply be summed up:
\begin{equation}
  H^i(E) = \sum_{j=1}^p H^i_j(E)
\end{equation}
Afterwards, the consecutive weights are calculated from the previous weights together with the total histogram
according to whatever rule one may choose, in our case by a recursive scheme~\cite{MUCA3}. 

Altogether, this leads to an efficient parallelization due to a very limited amount of communication.
Below, we will show the applicability of the method to a general problem, here the condensation of a lattice 
gas in two and three dimensions. 
As in real life and other than in~\cite{PMUCA}, we will not pay attention to the optimization of each degree
of parallelization but want to demonstrate the value of this parallelization to modern computational 
statistical physics.

\section{Condensation of a lattice gas}
\begin{figure*}
  \includegraphics[width=0.45\textwidth]{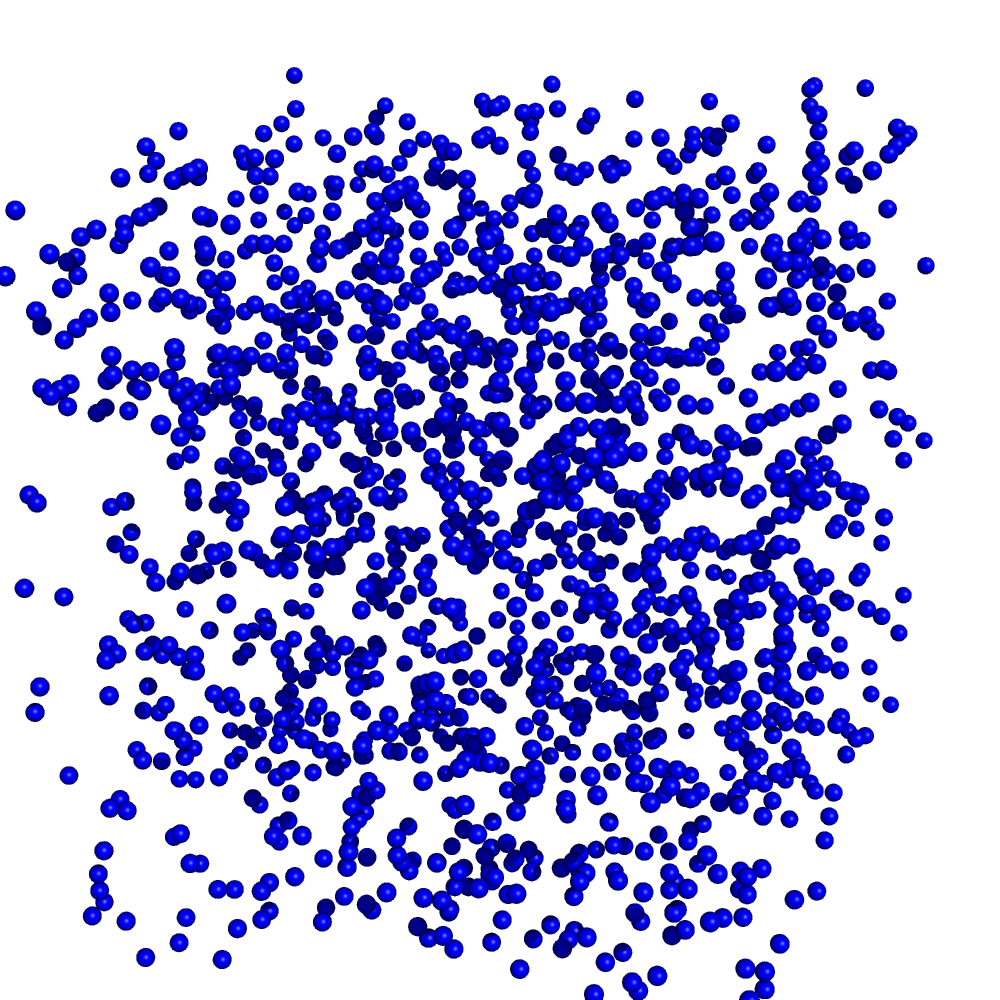}
  \hfill
  \includegraphics[width=0.45\textwidth]{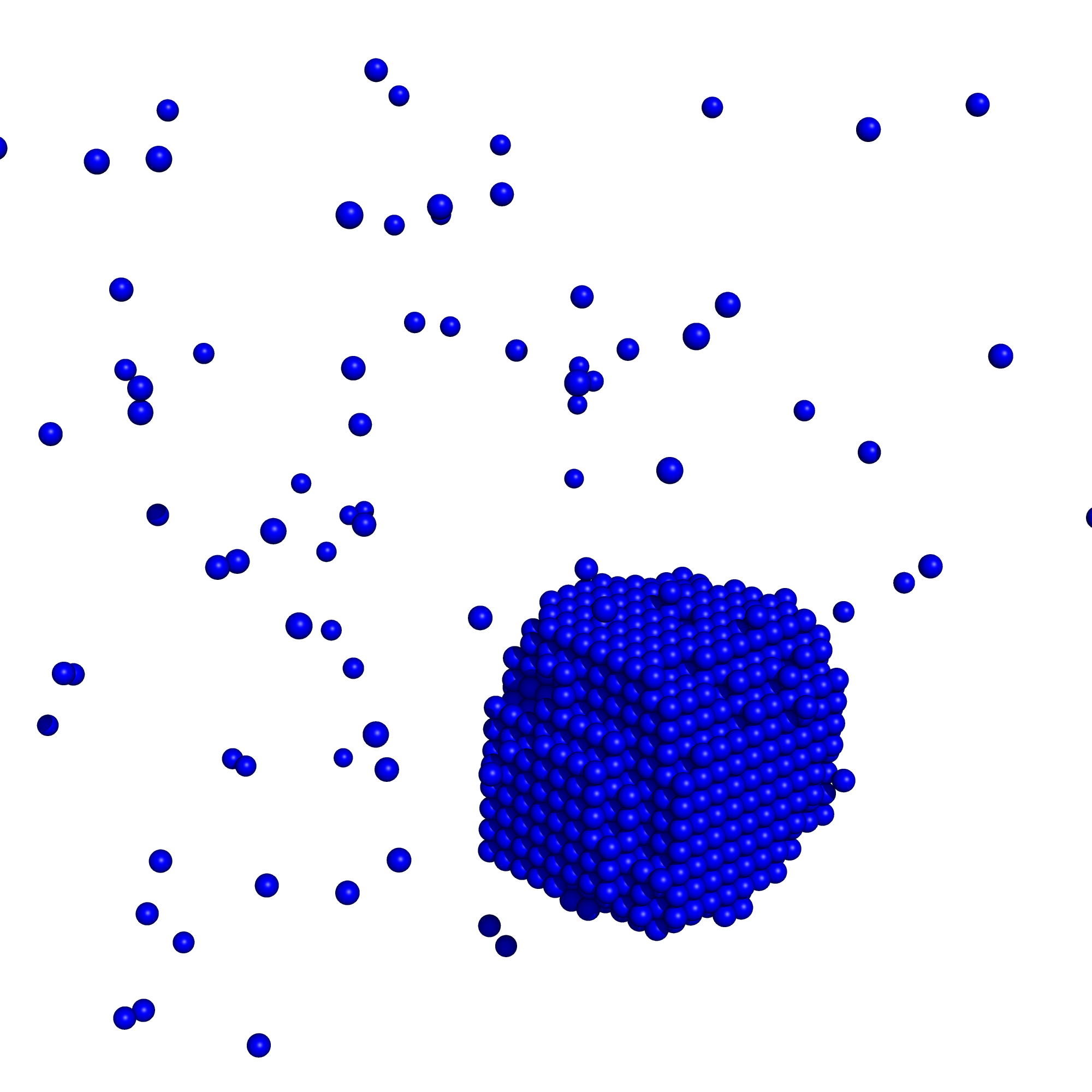}
  \caption{\label{fig:gas}
           Example of 
           (left) fluctuations of particle excess and 
           (right) a single condensate 
           in three dimensions.}
\end{figure*}

A general theory of equilibrium droplet condensation was proposed by 
Biskup et  al.~\cite{Biskup1,Biskup2}, with a rigorous analytic solution for the
two-dimensional Ising spin model. In principle they describe the interplay
of entropy maximization by vacuum fluctuations with energy minimization by
forming a droplet. They showed that the probability for intermediate-size droplets vanishes with 
increasing system size. 
The free energy consists of a contribution from the vacuum fluctuation of excess particles $\delta N$ 
\begin{equation}
  F_{\rm fluc} = \frac{(\delta N)^2}{2\hat{\kappa} V},
\end{equation}
with the isothermal compressibility 
\mbox{$\hat{\kappa}=\beta\kappa=\beta\left\langle\left(N-\langle N\rangle\right)^2\right\rangle/V$} 
and a contribution from the single largest droplet of size $V_D$
\begin{equation}
  F_{\rm drop} = \tau_W(V_D)^\frac{d-1}{d},
\end{equation}
where $\tau_W$ is the surface free energy of a (Wulff shaped) droplet of unit volume
(see Fig.~\ref{fig:gas}). 

This was translated into the language of the Ising spin model in \cite{Biskup1, Nussbaumer1}. 
Here, we will concentrate on the formulation of a lattice gas, where each lattice site is
either occupied by a particle or empty, i.e. $n_i\in\{0,1\}$. 
For a fixed number of particles this model is equivalent to the Ising model with spins 
$s_i\in\{-1,1\}$, at fixed total magnetization, when $s_i=2n_i-1$:
\begin{equation}
  \mathcal{H}^{\rm gas} = -J \sum_{<i,j>} n_i n_j =  \frac{1}{4}\mathcal{H}^{\rm Ising} + c(N)
\end{equation}
Thus, for fixed coupling constants, a scaling of the temperature by $T^{\rm Ising}=T^I=4T$ and the addition of 
a system-size dependent constant transforms the energy scale of an Ising model into the corresponding one of a lattice gas. 
All other observables are then mapped accordingly by scaling the temperature.

The contributions to the free energy can as well be formulated in the language of a lattice gas.
Consider a temperature-dependent vacuum or background density $\rho_0=N_0/V$, which may be 
identified with the spontaneous magnetization in the Ising model via 
\begin{equation}
  m_0=1-2\rho_0.
\end{equation}
Adding particles to the system leads to a total particle excess $\delta N = N-N_0$.
According to Biskup et  al.\ this excess $\delta N$ may be decomposed into 
the particle excess inside the droplet $\delta N_D$ and the remaining particle excess in the 
fluctuating phase $\delta N_F$.
For the infinite-size system and from the equivalence to the Ising model follows that the
gas density is given by $\rho_G=\rho_0$ and the liquid density by $\rho_L=1-\rho_0$.
Hence, in a droplet of size $V_D$ we expect an excess of particles \mbox{$\delta N_D=(\rho_L -\rho_G)V_D$}.
Introducing the fraction of excess particles inside a droplet \mbox{$\lambda=\delta N_D/\delta N$}, allows 
us to rewrite the decomposed particle excess as fractions $\delta N_D=\lambda \delta N$ and 
$\delta N_F = (1-\lambda)\delta N$. 
Altogether the total free energy $F=F_{\rm drop}+F_{\rm fluc}$ then reads 
\begin{equation}
  F = \left(\tau_W \left( \frac{\lambda\delta N}{1-2\rho_0}\right)^{\frac{d-1}{d}} + \frac{(1-\lambda)^2(\delta N)^2}{2\hat{\kappa} V} \right).
\end{equation}
Rewriting the particle excess in terms of a volume 
\begin{equation}
  V_L = \frac{N-N_0}{1-2\rho_0}
\end{equation}
and identifying $\lambda=V_D/V_L$ leads to
\begin{eqnarray}
  F &=& \left(\tau_W (\lambda V_L)^{\frac{d-1}{d}} + \frac{(1-2\rho_0)^2}{2\hat{\kappa} V}V_L^2(1-\lambda)^2 \right) \nonumber\\
  &=& \tau_W V_L^{\frac{d-1}{d}}\left(\lambda^{\frac{d-1}{d}} + \Delta (1-\lambda)^2\right) \label{eq:freeEnergy}
\end{eqnarray}
with the dimensionless parameter
\begin{eqnarray}\label{eq:Delta}
  \Delta &=& \frac{(1-2\rho_0)^2}{2\hat{\kappa} \tau_W} \frac{V_L^{(d+1)/d}}{V}. 
\end{eqnarray}
 
While $\hat{\kappa}$ has the same value as $\chi$ in the Ising model, $\tau_W$ has to be rescaled such that
$\tau^I_W=4\tau_W$. 
At fixed temperature, the parameters $\rho_0,\hat{\kappa},\tau_W$ are constant.
Minimizing eq.\ (\ref{eq:freeEnergy}) leads to the infinite-size solution for the equilibrium volume fraction
$\lambda=\lambda(\Delta)$ (see~\cite{Biskup1,Biskup2}). Biskup et al.\ showed
that below a critical value $\Delta_c^{2D}\approx0.9186$ or $\Delta_c^{3D}\approx0.8399$ there are merely gas 
fluctuations, while above there exists one macroscopic droplet together with background fluctuations. 
At $\Delta_c$ the mass of the largest droplet jumps to a fraction $\lambda_c^{2D}=2/3$ or $\lambda_c^{3D}=1/2$ 
of the total particle excess.
Droplets of intermediate size have negligible probability.

In order to compare to the analytic predictions, we have to measure the average size of the largest droplet
at different densities but fixed temperature. Because of the increasing correlation times in large systems,
we applied the multicanonical method, driving the system between the evaporated and condensed phase,
and reweighted to the relevant temperatures afterwards. The energy ranges were obtained through short 
parallel tempering~\cite{PT1} runs in the beginning of each simulation.

\section{Results}
\begin{figure}
  \includegraphics[width=0.5\textwidth]{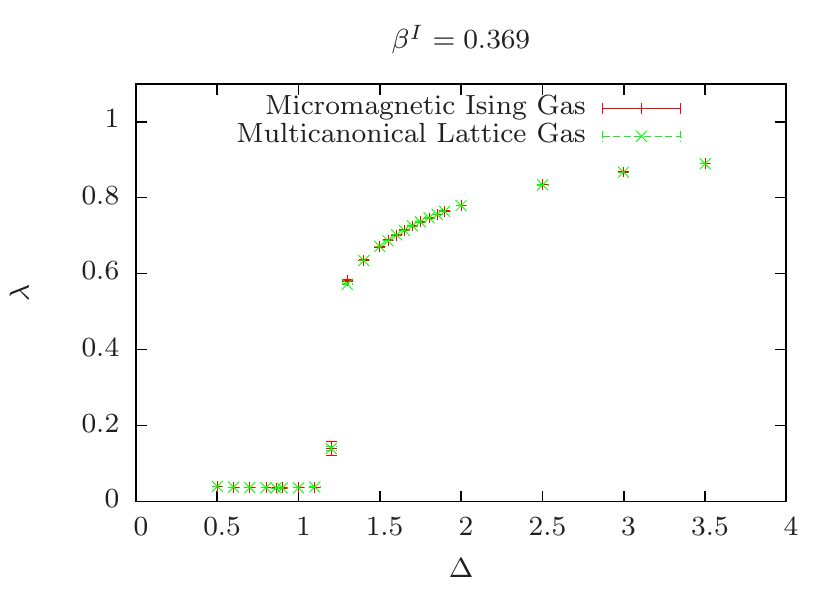}
  \begin{minipage}[b]{0.5\textwidth}
    \caption{
      \label{fig:compare}
      Comparison of simulation data from micromagnetic Metropolis simulations of the Ising model and
      parallel multicanonical simulations of the lattice gas in three dimension. Here, the volume
      is fixed at $L=30$ while the number of spins/particles are modified. 
    }
    \vspace{4em}
\end{minipage}
\end{figure}

In order to assess the parallel algorithm, we performed two independent simulations of equivalent
but different models, namely the Ising spin model and the lattice gas. While the lattice gas was simulated
with the parallel multicanonical algorithm explained above, the Ising model was simulated with a 
Metropolis algorithm at fixed magnetization, also referred to as micromagnetic simulation.
Both approaches needed independent simulation runs for each measurement point.
We chose the inverse temperature in the Ising language to be $\beta^I=0.369$, which corresponds to a 
temperature $T=0.6995$ in the particle picture.
Figure~\ref{fig:compare} shows very good agreement between the two completely independent and 
different methods.

A direct comparison of simulation times is rather difficult. For one, the multicanonical method
provides additional information for various temperatures, and this comes at a price: 
A time-consuming weight iteration, which is not contributing to the production run. 
Whether or not the overhead of the initial weight iteration is justified, will depend on the 
desired size of the statistical error.
Furthermore, the approaches will scale differently with system size, which we do not want to address at 
this point.
For the plot in Fig.~\ref{fig:compare}, the average simulation time per data point of both methods
was about half an hour on common 2.5GHz cores, but the multicanonical simulation used 36 cores instead
of one in the case of the micromagnetic simulation. 
However, with increasing system size the free-energy barrier between the evaporated and the condensed 
phase increases \cite{Nussbaumer3}, and the necessary computation time for the
single-core micromagnetic simulation exceeds the limits of practicability.

Next, we discuss the speedup of the parallel method with increasing degree of parallelization. 
As mentioned above, we intentionally did not pay attention to optimization of the simulation parameters.
Rather we want to demonstrate the applicability of the parallelization by choosing appropriate parameters
from single-process simulations.
Hence, we fixed the total number of sweeps per iteration to $25600$ such that each process performed
$25600/p$ sweeps. A single sweep consists of $L^d$ updates. The number of particles was chosen to be
$N=1000$ and the energy range adapted accordingly. The simulation was defined to be converged if a total
of 30 ``tunneling events'' were counted, meaning that the simulations had traveled at least 30 times 
from the maximal energy to the minimal energy or vice versa.

\begin{figure}
  \includegraphics[width=0.5\textwidth]{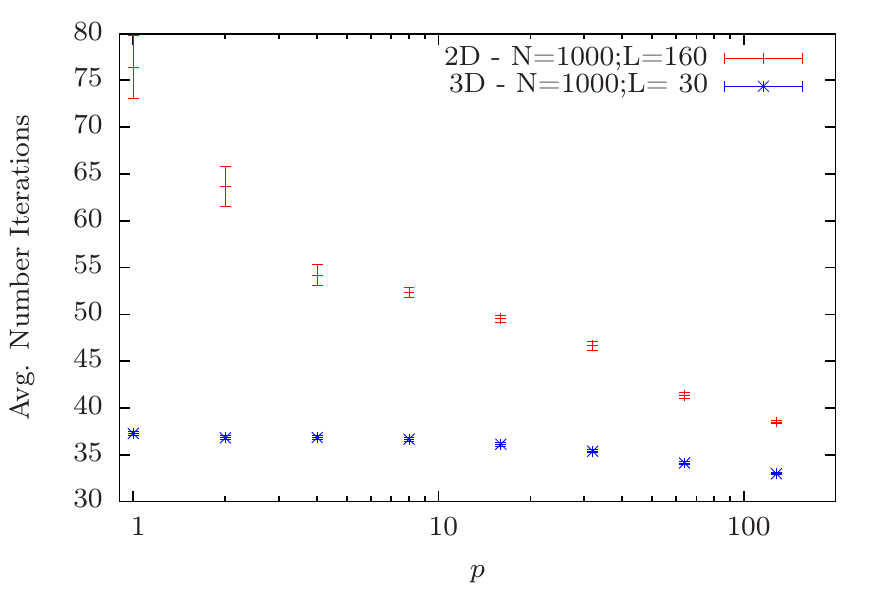}
  \includegraphics[width=0.5\textwidth]{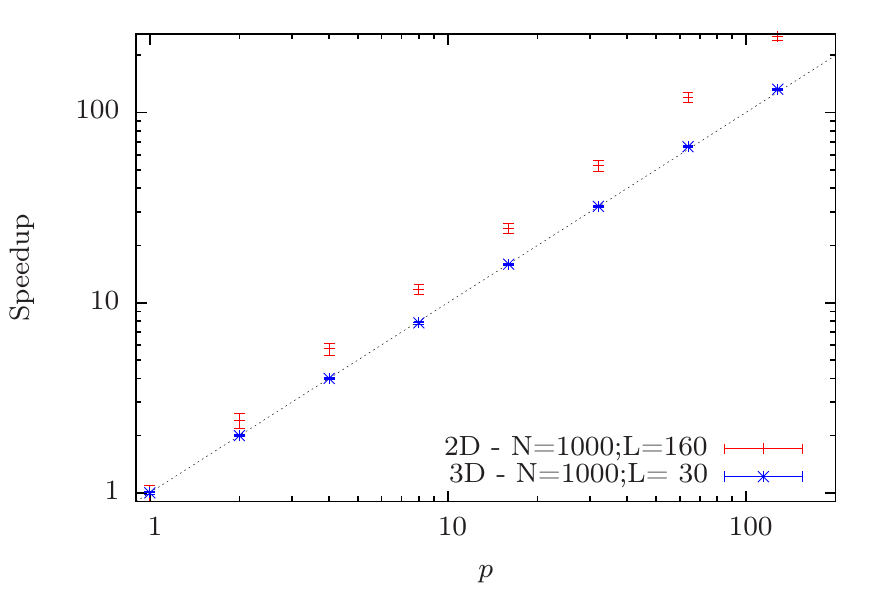}
  \caption{
    \label{fig:scaling}
    (a) Average number of iterations until convergence.
    (b) Average time speedup. The dotted line is a linear function of slope 1.
    Each average consists of 32 parallel multicanonical simulations.
  }
\end{figure}

When we want to estimate the speedup, we need to consider statistical averages.
This is due to the fact that the weight iteration relies on the sampled data such that the number of iterations
until convergence cannot be predicted and may, moreover, vary depending on the initial seed of the 
random number generator.
To this end, we performed 32 simulations for each degree of parallelization.
The effect is best seen in the average number of iterations until convergence (\mbox{Fig.~\ref{fig:scaling}~(a)}). 
An increasing number of independent Markov chains improves each estimate of the distribution belonging 
to the current weights. 
This optimizes the estimation for the successive weights and reduces the total amount of necessary number
of iterations. 
Of course, this is in general not true on all scales~\cite{PMUCA}, for example when the number of sweeps
per iteration becomes too small or if the optimal number of sweeps is more complex than the assumed
$1/p$ behavior. 

The speedup is defined in terms of the average time until convergence; it is given by the ratio of a 
single-core to a $p$-core simulation: 
\begin{equation}
  S_p = \frac{\bar{t}_1}{\bar{t}_p}.
\end{equation}
For a fair comparison, both times were obtained with the same program.
Because the communication between processes occurs only at the end of every iteration, the speedup should be
ideal with a linear behavior of slope one, i.e., doubling the amount of processes should speed up the time by a
factor 2.
We can see in \mbox{Fig.~\ref{fig:scaling}~(b)} that this is indeed true for the case of three dimensions and
even better for two dimensions. This may be explained by the reduced number of iterations, due to the 
independence of Markov chains.

\begin{figure}
  \includegraphics[width=0.5\textwidth]{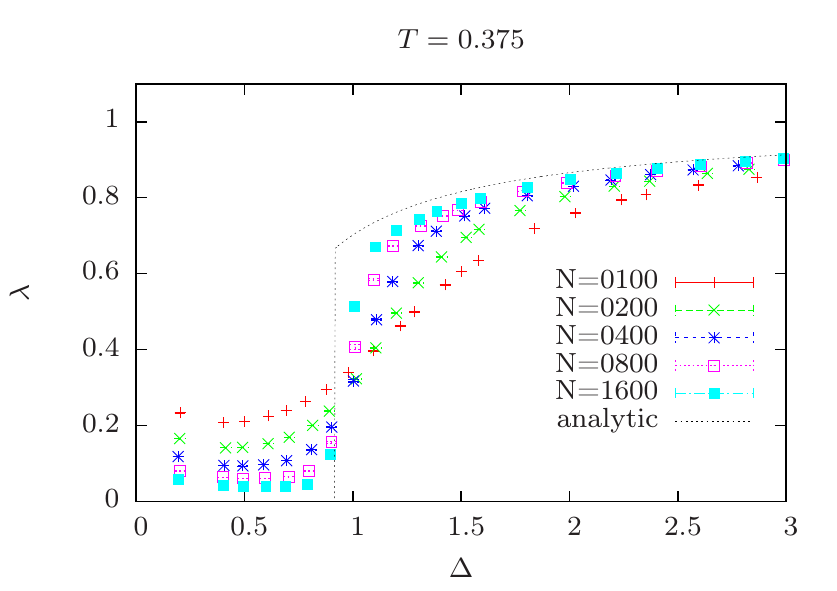}
  \includegraphics[width=0.5\textwidth]{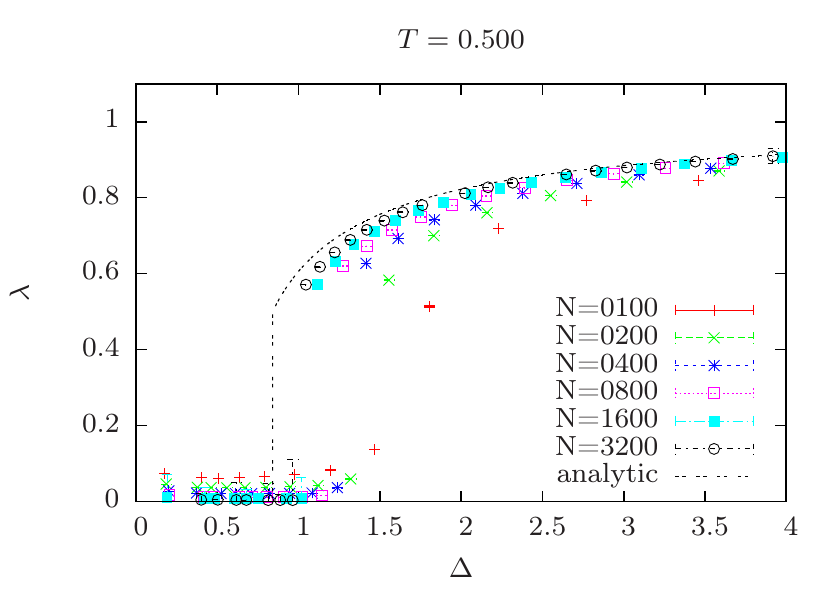}
  \caption{
    Condensation of a lattice gas in two (left) and three (right) dimensions. The temperature is chosen, such 
    that the two-dimensional case corresponds to the Ising model at $T^I=1.5$ in Refs.~\cite{Nussbaumer1,Nussbaumer2}
    and the three-dimensional case is below the roughening transition.
    The differently colored curves belong to fixed number of particles $N$, while the density is varied by 
    changing $L$. Each data point corresponds to an independent simulation with up to $24$ cores.
  }
  \label{fig:condensation}
\end{figure}

Having demonstrated the correctness of our model and the efficiency of our method, we want to apply it
to the problem of condensation in a lattice gas.
Figure~\ref{fig:condensation} shows the results for selected temperatures in two and tree dimensions
with up to $24$ cores in each simulation. 
The temperature in two dimension was chosen to correspond to the equivalent Ising temperature \mbox{$T^I=1.5$}
used in Ref.~\cite{Nussbaumer1}, allowing us to use the same parameters.
In three dimension we chose a temperature $T^I=2$ below the roughening transition 
$T^I_{R}\approx2.4537$~\cite{TR}, where low-temperature approximations hold. 
This allowed us to use a low-temperature series expansion for the interface tension: 
$\sigma=1.9072\dots$~\cite{Tau_Low}.
Furthermore, below the roughening transition the condensate tends to form a cube, such that the surface 
free energy may be assumed as $\tau^I_{W}\approx6\sigma$.
This is the result for the Ising model and has to be rescaled by the same factor as the temperature,
hence $\tau_W\approx6\sigma/4$ (the same argument applies to the two-dimensional case).
The remaining free parameters, $\rho_0=(1-m_0)/2$ and $\hat{\kappa}=\chi$, were obtained by 
standard Metropolis simulations of the Ising model on very large lattices.
An overview of the involved parameters is given in Table~\ref{tab:parameter}.
\begin{table}[b]
  \caption{Converted parameters for the lattice gas system in two~\cite{Nussbaumer2} and three dimensions, where 
    for the latter case \mbox{$\rho_0=(1-m_0)/2$} and $\hat{\kappa}=\chi$ are obtained from Metropolis simulations of the Ising 
    model at various system sizes and $\tau_W\approx6\sigma/4$ with $\sigma$ from a low-temperature 
    expansion~\cite{Tau_Low}.}
  \label{tab:parameter}
  \centering
  \begin{tabular}{|l|c|c|}
    \hline
    Parameter       & 2D                    & 3D          \\
    \hline  
    $T$             & 0.375                 & 0.500          \\
    $\rho_0$        & 0.00675               & 0.002739(1)    \\ 
    $\hat{\kappa}$  & 0.02708               & 0.00608(1)     \\
    $\tau_W$        & 1.06125               & $\approx2.861$ \\
    \hline
  \end{tabular}
\end{table}

Opposite to Ref.~\cite{Nussbaumer1}, we always fixed the number of particles and varied the
density by modifying the volume of the system.
We can see in Fig.~\ref{fig:condensation} that this choice leads to a similar behavior in two dimensions 
as in Ref.~\cite{Nussbaumer1}, where the predicted $\Delta_c$ separates the evaporated from
the condensed phase quite well. Of course, we see finite-size effects as reported before, but the qualitative
picture is satisfying.
In three dimensions, on the other hand, we observe a strong deviation of small systems from the
predicted critical value of the parameter $\Delta$. 
This becomes weaker but is still present for the largest systems investigated. 
Specifically, the system needs to go to larger $\Delta$ or, equivalently, particle density in order to
condensate. 
With increasing system size, the critical value of $\Delta$ decreases, while the shape of the curves seems to 
converge to the predicted functional shape.
This may be a finite-size effect that is more pronounced in three dimensions, which calls for more detailed 
studies in future work.
In both cases, we do not have any free parameters left.

\section{Conclusion}
We have demonstrated the applicability of an efficient but simple parallelization of the multicanonical method
to the case of lattice gas condensation in two and three dimensions.
In these cases, the speedup was shown to scale ideally (or even better) due to the fact that 
independent Markov chains improve the sampling of equilibrium distributions.

Applying this method to the problem of condensation, we could reproduce the results from~\cite{Nussbaumer1} 
in two dimensions and present new results in three dimensions which indicate large finite-size effects. 
In the latter case, condensation systematically occurred at densities higher than predicted. 
While the deviation decreases with system size, the available data does not allow for quantitative 
predictions due to the low resolution at the critical density caused by varying the volume.
This has still to be investigated in more detail.

\section*{Acknowledgments}
We would like to thank Martin Marenz for the joint development of a Monte Carlo Simulation Framework.
This work has been partially supported by
  the Leipzig Graduate School of Excellence GSC185 ``BuildMoNa'',
  the Deutsch-Franz\"osische Hochschule DFH-UFA (grant CDFA-02-07) 
  and the S\"achsische DFG-Forschergruppe FOR877.
The project was funded by the European Union and the Free State of Saxony.

\section*{References}
\bibliography{zierenberg.bib}

\end{document}